\documentclass[usenatbib,fleqn,twocolumn]{mnras}
\usepackage{graphicx}
\usepackage{amsmath}
\usepackage{amssymb}
\usepackage{bm}
\usepackage{hyperref}
\usepackage{natbib}
\usepackage{color}
\usepackage[T1]{fontenc}

\newcommand{\beq}{\begin{equation}}
\newcommand{\eeq}{\end{equation}}

\newcommand{\benum}{\begin{enumerate}}
\newcommand{\eenum}{\end{enumerate}}

\def\apj{ApJ}

\def\aap{A \& Ap}

\def\pasj{PASJ}

\def\araa{ARAA}

\def\mnras{MNRAS}
\def\apj{ApJ}
\def\apjl{ApJL}
\def\aap{A\&A}

\def\physrep{Physics Reports}
\def\pre{Physical Review E}
\def\physscr{Physica Scripta}

\def\prl{Phys. Rev. Lett.}

\newcommand{\abra}[1]{\left\langle{#1}\right\rangle}

\newcommand{\del}{{\bm\nabla}}
\newcommand{\bmA}{{\bm A}}
\newcommand{\bma}{{\bm a}}
\newcommand{\bmB}{{\bm B}}
\newcommand{\bmb}{{\bm b}}

\newcommand{\bmj}{{\bm j}}

\newcommand{\bmU}{{\bm U}}
\def\bmu{\bm{u}}

\newcommand{\emf}{\mathcal{E}}

\newcommand{\alphaM}{\alpha^\text{m}}

\newcommand{\alphaSS}{\alpha_\text{SS}}
\newcommand{\cs}{c_\text{s}}
\newcommand{\cso}{c_\text{s0}}

\newcommand{\rin}{r_\text{in}}

\newcommand{\Sh}{\text{Sh}}
\newcommand{\betaL}{\beta_\text{L}}
\newcommand{\betaLeq}{\beta_\text{L,eq}}
\newcommand{\betaLst}{\beta_\text{L,st}}
\newcommand{\run}[1]{run~\texttt{#1}}
\newcommand{\fw}{f_\text{w}}
\newcommand{\ff}{f_\text{f}}

\begin{document}

\title{Helical and nonhelical large-scale dynamos in thin accretion discs}
\author[Zhou]{Hongzhe Zhou$^{1}$%
\thanks{Email address for correspondence: hongzhe.zhou@sjtu.edu.cn}\\
$^1$
Tsung-Dao Lee Institute, Shanghai Jiao Tong University,
800 Dongchuan Road, Shanghai 200240,
People's Republic of China
}

\date{\today}
\maketitle

\begin{abstract}
The dynamics of accreting and outgoing flows around compact objects depends crucially on the strengths and configurations of the magnetic fields therein,
especially of the large-scale fields that remain coherent beyond turbulence scales.
Possible origins of these large-scale magnetic fields include flux advection and disc dynamo actions.
However, most numerical simulations have to adopt an initially strong large-scale field rather than allow them to be self-consistently advected or amplified, due to limited computational resources.
The situation can be partially cured by using sub-grid models where dynamo actions only reachable at high resolutions are mimicked by artificial terms in low-resolution simulations.
In this work, we couple thin-disc models with local shearing-box simulation results to facilitate more realistic sub-grid dynamo implementations.
For helical dynamos, detailed spatial profiles of dynamo drivers inferred from local simulations are used, and the nonlinear quenching and saturation is constrained by magnetic helicity evolution.
In the inner disc region, saturated fields have dipole configurations and the plasma $\beta$ reaches $\simeq 0.1$ to $100$,
with correlation lengths $\simeq h$ in the vertical direction and $\simeq 10h$ in the radial direction, where $h$ is the disc scale height.
The dynamo cycle period is $\simeq 40$ orbital time scale, compatible with previous global simulations.
Additionally, we explore two dynamo mechanisms which do not require a net kinetic helicity and have only been studied in shearing-box setups.
We show that such dynamos are possible in thin accretion discs, but produce field configurations that are incompatible with previous results.
We discuss implications for future general-relativistic magnetohydrodynamics simulations.
\end{abstract}


\section{Introduction}
Accretion discs around compact objects power the most luminous objects in the Universe; yet, a thorough understanding of how the gravitational energy of the in-falling matter is converted into radiation is still lacking.
In recent years, a common consensus has been built that magnetic fields play central roles in the dynamics of accretion discs, including initiating turbulent flows, enhancing angular momentum transport, and launching jets \citep{BalbusHawley1991,Narayan+2003,YuanNarayan2014}.
In cases where the central engine is a black hole, jet launching through the Blandford-Znajek mechanism depends on the net magnetic flux accumulated on the horizon \citep{Tchekhovskoy+2011}.

General-relativistic magnetohydrodynamics (GRMHD) simulations have become increasingly feasible to study accretion physics in the vicinity of black holes.
Given that it is still costly to resolve turbulent flows to the extent where the magnetic field amplification therein, i.e., the dynamo processes, becomes resolution-independent,
it has been a common practice to start simulations with gas already threaded by some large-scale poloidal magnetic fluxes.
Yet, simulation results rely on how such initial magnetic fields are arranged \citep[e.g.,][]{Beckwith+2008,BarkovBaushev2011,Narayan+2012}:
single-loop poloidal fields lead to strong net fluxes on the horizon and therefore more powerful jets,
and fields with alternating polarity lead to episodic and less powerful jets.
It remains an active research topic how such large-scale fields should be properly modeled and how the accretion dynamics is affected.

There are two possible origins for the large-scale fields in accretion discs: 
(i) the large-scale fields are amplified \textit{ab initio} in the accretion flow, i.e., through large-scale dynamo (LSD) actions, and
(ii) such large-scale fields are advected from the outer disc region.
For the latter scenario to work, the diffusion time scale of the large-scale fields has to be longer than the advection time scale.
Shearing-box simulations suggest that the turbulent magnetic Prandtl number is of order unity \citep{GuanGammie2009}, seemingly implying inefficient advection in thin discs, but the corona field \citep{Beckwith+2009}, disc vertical structure \citep{GuiletOgilvie2012,GuiletOgilvie2013}, or disc winds \citep{CaoSpruit2013} may help alleviate the problem.
For thick discs, advection can be more efficient \citep{Cao2011,Ressler+2020,Dhang+2023apj}.

In this work, we focus on the alternative explanation of disc magnetization, i.e., through LSDs.
\textit{Ab initio} amplification of magnetic fields that are coherent above the turbulence scale have been observed in both local shearing-box \citep{Brandenburg+1995,BaiStone2013} and global Newtonian or post-Newtonian \citep{HoggReynolds2016,RodmanReynolds2023} simulations, but much less often reported in the costly GRMHD simulations \citep{Liska+2020}.

Mean-field electrodynamics is commonly used to understand the dynamics of large-scale magnetic fields in plasmas.
In such a framework, turbulent fields are averaged out and their mean dynamics contribute to the evolution of the mean magnetic field \citep[for a comprehensive review, see][]{BrandenburgSubramanian2005}.
The coherent effect on the mean magnetic field from random turbulent fields is manifested as the turbulent electromotive force (EMF), $\bm\emf$, in the induction equation.
Mathematically, $\bm\emf=\abra{\bmu\times\bmb}$, where $\abra{\cdots}$ means an ensemble average and $\bmu$ and $\bmb$ are turbulent velocity and magnetic fields, respectively.
A common practice is to expand $\bm\emf$ in terms of the gradients of the mean magnetic field $\bmB$,
\beq
\emf_i=\alpha_{ij}B_i+\eta_{ijk}\partial_j B_k+\cdots,
\eeq
where $\alpha_{ij}$ and $\eta_{ijk}$ are tensorial turbulent transport coefficients.
A key question is then how these dynamo coefficients can be written in terms of other mean quantities in the system, such as density stratification and differential rotation.
A direct way is to measure these turbulent transport coefficients in numerical simulations, and sophisticated tools have been invented to do the job.
The test-field method \citep{Brandenburg+2008,RheinhardtBrandenburg2010,Kapyla+2022}
has been applied in local MRI simulations \citep{Gressel2010},
and the projection method \citep{BrandenburgSokoloff2002} can be applied to more complicated geometries, such as in thick-disc simulations  \citep{Dhang+2020}.

One of the practical purposes of a comprehensive understanding of dynamo processes is for better sub-grid dynamo models.
Three-dimensional (3D) simulations of accretion discs are capable of dynamos,
but are computationally expensive to achieve a resolution-independent regime.
In contrast, two-dimensional (2D) simulations are less costly, but intrinsically cannot host a dynamo \citep{Cowling1933}.
In order to mimic dynamo processes that are available only at high resolution, sub-grid dynamo models are used in 2D or low-resolution 3D simulations, where a turbulent EMF is artificially added to the induction equation to amplify magnetic fields.

The implementation and application of helical sub-grid dynamo effects have been explored by several groups \citep{vonRekowski+2003,BucciantiniDelZanna2013,Bugli+2014,StepanovsFendt2014,Skadowski+2015,FendtGassmann2018,Dyda+2018,Tomei+2020,VourellisFendt2021}.
However, previous sub-grid models have not exploited the results from analytical theories and local shearing-box simulations:
a shortcoming of these works is that both the nonlinear feedback from the magnetic field to the $\alpha_{ij}$ coefficient (i.e., dynamo quenching) and the saturation strength employ parameterised forms.
In the present work, we improve the previous helical dynamo models in the following ways:
(i) the spatial profile of $\alpha_{ij}$ is derived based on previous analytical and test-field simulations results, and
(ii) the quenching prescription of $\alpha_{ij}$ and saturation level of the large-scale field are determined by magnetic helicity evolution.
As we shall see, the new model yields dynamo waves and cycle periods in agreement with both local and global direct numerical simulations (DNS) of accretion-disc dynamos.

In addition to the helical dynamo effect, a new class of shear dynamos has been noticed in shearing-box simulations \citep{Brandenburg2005,Yousef+2008},
where \emph{nonhelical} turbulence and a background shear flow are the only ingredients necessary for the growth of a large-scale magnetic field.
Due to such rather simple requirement, it is expected that shear dynamos may widely operate in differentially rotating flows, such as accretion discs.
How shear dynamos should be modeled is still under debate \citep{ZhouBlackman2021shear}, and two theoretical models have been proposed:
(i) the shear-current effect \citep[SCE;][]{RogachevskiiKleeorin2003,SquireBhattacharjee2015}, and
(ii) the incoherent-$\alpha$ effect \citep[IAE;][]{VishniacBrandenburg1997,SridharSingh2014,Jingade+2018}.
Both theories have only been demonstrated in shearing-box setups.
In this work, we implement these nonhelical dynamo mechanisms in sub-grid models,
and investigate their dynamo capabilities in the disk geometry.

The rest of this work is organized as follows.
In section~\ref{sec:model} we introduce the axisymmetric model which couples a thin disc with helical or nonhelical large-scale dynamos.
In section~\ref{sec:simulations} we introduce the simulation setup that numerically solves the derived models, and the results are presented and discussed in section~\ref{sec:results}.
We conclude in section~\ref{sec:conclusion}.

\section{Axisymmetric disc-dynamo model}
\label{sec:model}
In this section we introduce the mean-field model that applies to a geometrically thin, optically thick disc.
We consider a fully turbulent disc which results from the magneto-rotational instability 
\citep[MRI;][]{Velikhov1959,Chandrasekhar1961,BalbusHawley1991},
whose turbulent time scale is comparable to the local Keplerian time scale and hence much shorter than that of the LSD.
Focusing on the LSD processes, we neglect the back-reaction of the LSD-generated magnetic field onto the turbulence and disc structure in this work.
In section~\ref{sec:conclusion} we discuss the possible outcomes of such feedbacks and the possibility of a more complete disc-dynamo model.

Throughout this work, the physical quantities mentioned (e.g., gas density, pressure, magnetic fields, etc.) all refer to the large-scale ones, i.e., those survive after averaging out the turbulent fluctuations, unless otherwise specified.
We use $(r,\theta,\phi)$ for spherical coordinates,
and $(\varpi,\phi,z)$ for cylindrical coordinates.

\subsection{The thin accretion disc}
We consider a Shakura-Sunyaev type disc filled with polytropic gas,
\beq
P=P_0\left(\frac{\rho}{\rho_0}\right)^\gamma,
\eeq
where $P$ is the gas pressure, $\rho$ is the density, $\gamma$ is the ratio of specific heats, and a subscript $0$ indicates quantities measured at the disc mid-plane at $r=r_0$.
The thermal sound speed is
\beq
\cs=\sqrt{P/\rho},
\eeq
and hence $P_0=\rho_0\cso^2$.
The vertical balance between the gravitational force and the gas pressure gradient near the disc mid-plane gives
\beq
h=\cs/\Omega,
\label{eqn:h}
\eeq
where $\Omega$ is the Keplerian rotation rate.
We denote $\epsilon=h/\varpi$ as the dimensionless disc scale height, and define $|\theta-\pi/2|\geq 2\epsilon$ as the corona region.
$h$ is also taken to be the typical scale of the growing mode of the LSDs, and the corresponding wave number is $k_1=h^{-1}$.

The turbulence time scale is roughly $\tau=\Omega^{-1}$.
We denote the turbulence length scale by $l$, which also gives the turbulence velocity scale $u=l/\tau=l\Omega$.
The typical wave number of turbulent fluctuations is then $k_2=l^{-1}$.

The turbulent viscosity is
\beq
\nu=\alphaSS\cs h,
\label{eqn:nu}
\eeq
where $\alphaSS$ is the Shakura-Sunyaev parameter.
The turbulent viscosity should be of order $\nu\sim ul=l^2\Omega$.
Combining with equations~(\ref{eqn:h}) and (\ref{eqn:nu}), we have the estimates \citep{Blackman1998}
\beq
l\simeq \sqrt{\nu/\Omega}=\sqrt{\alphaSS\cs h/\Omega}
=\alphaSS^{1/2}h,
\label{eqn:l}
\eeq
and
\beq
u=l\Omega\simeq\alphaSS^{1/2}\cs,
\label{eqn:u}
\eeq
which are useful when deriving dynamo coefficients in terms of $\cs$ and $\Omega$.

We ignore the feedback from large-scale fields on the turbulence, and use $\alphaSS=0.3$ \citep{King+2007} which remains constant and uniform.

\subsection{Mean-field dynamo terms}
The evolution equation for the large-scale magnetic field can be derived by averaging the induction equation over a suitable scale that is much larger than the turbulence scale.
The mean-field equation is \citep[for a derivation see, e.g.,][]{BrandenburgSubramanian2005}
\beq
\partial_t\bmB=\del\times\bm\emf+\del\times\left(\bm U\times\bm B\right)+
\del\times\left[\left(\bm\Omega\times\bm\varpi\right)\times\bm B\right],
\label{eqn:induc}
\eeq
where $\bm B$ is the large-scale magnetic field,
$\bm U$ is the mean flow including the accretion inflow and possible large-scale outflows like disc winds, but excluding the Keplerian motion,
and $\bm\emf$ is the turbulent EMF.

The last term in equation~(\ref{eqn:induc}) generates toroidal fields from the poloidal ones through differential rotation (the $\Omega$ effect), and hence for a successful dynamo,
the turbulent EMF must contain terms that generates poloidal fields from the toroidal ones.
In this work we consider the turbulent EMFs of the form
\beq
\emf_i=\alpha_{ij}B_j-\eta_{ij}J_j+\cdots,
\label{eqn:emf}
\eeq
where $\alpha_{ij}$ and $\eta_{ij}$ are turbulent transport coefficients,
and $\bm J=\del\times\bm B/\mu_0$ is the mean current density.

The diagonal components of $\eta_{ij}$ contribute to the turbulent diffusion of large-scale magnetic fields.
We consider isotropic diffusion, and the diagonal components of $\eta_{ij}$ are
\beq
\eta_{rr}=\eta_{\theta\theta}=\eta_{\phi\phi}=
\eta=\alphaSS\cs h\left[\frac{3}{8}+\frac{3}{8}\left(\frac{z}{h}\right)^2\right].
\label{eqn:eta}
\eeq
Here the first factor, $\alphaSS\cs h$, is the same as the turbulent viscosity, and the second factor in the square brackets phenomenologically captures the vertical structure of $\eta$, inspired by test-field results in stratified shearing-box simulations \citep{Gressel2010}.
The factor of $3/8$ ensures that the average of $\eta$ over one disc scale height is $\alphaSS\cs h$, and hence we consider a turbulent magnetic Prandtl number of order unity.
The coupling of large-scale fields and turbulence will lead to modifications of $\eta$, which we do not include in this work.
Some dynamo models using parametrized forms of $\eta$ include \cite{Zanni+2007} and \cite{StepanovsFendt2014a}.

The remaining components in $\alpha_{ij}$ and $\eta_{ij}$ vary in different models and are derived in the next few subsections.
In the follows, we describe in detail three dynamo models:
(i) the $\alpha\Omega$ dynamo,
(ii) the incoherent $\alpha$ dynamo, and
(iii) the shear-current dynamo.
The $\alpha\Omega$ model is the conventional helical dynamo that has been widely applied to explain stellar and galactic magnetization,
whereas the latter two need no net kinetic helicity and are less explored, particularly in a thin-disc geometry.
For each, we derive explicit forms of the turbulent EMF (\ref{eqn:emf}) with physics- or simulation-motivated quenching formulae, expressed in terms of $\cs$ and $\Omega$ that can be readily implemented in an accretion disc context.
A summary of dynamo models is in table~\ref{tab:dynamo},
and numerical solutions of the models are given in the next section.

\begin{table}
\centering
\caption{Summary of dynamo models.}
\label{tab:dynamo}
\begin{tabular}{cc}
\hline
Driver & EMF with quenching \\
\hline
$\alpha\Omega$ & (\ref{eqn:emf_ao}), (\ref{eqn:alphaij_ao}), (\ref{eqn:alpha_betaL})\\
incoherent-$\alpha$ & (\ref{eqn:emf_fa}), (\ref{eqn:alphatot_fa}) \\
shear-current & (\ref{eqn:emf_sc}), (\ref{eqn:eta_sc}), (\ref{eqn:Sh_sc}) \\
\hline
\end{tabular}
\end{table}

\subsection{The $\alpha\Omega$ model}
We consider a minimalist helical dynamo model where the off-diagonal components of $\eta_{ij}$ vanish, and hence the turbulent EMF takes the form
\beq
\emf_i=\alpha_{ij}B_j-\eta J_i.
\label{eqn:emf_ao}
\eeq
From stratified shearing-box simulations, all components of $\alpha_{ij}$ have roughly the same order of magnitude \citep{Gressel2010}, but the $\alpha_{\phi\phi}$ component is the most relevant as it is responsible for generating poloidal magnetic fields from toroidal ones.
Hence we only consider $\alpha_{\phi\phi}$ in what follows, making the model an $\alpha\Omega$-type dynamo.

It has been long-debated the profile of $\alpha_{\phi\phi}$ in accretion discs,
especially about how its sign changes with latitude.
In hydrodynamical cases, stratified rotating turbulence possesses a net kinetic helicity in each hemisphere, making the kinetic contribution to $\alpha$ positive in the upper plane and negative in the lower plane.
The picture changes in the presence of MRI, where local shearing-box \citep{Gressel2010,Dhang+2023} and global \citep{HoggReynolds2018,Dhang+2020} simulations report that $\alpha_{\phi\phi}$ has opposite signs to its hydrodynamic values within one to two disc scale heights.
This has been attributed to the combined effects of differential rotation and magnetic buoyancy \citep{Brandenburg1998},
or helicity flux \citep{GopalakrishnanSubramanian2023}.
Within one disc scale height, $\alpha_{\phi\phi}$ has the expected signs as in the hydrodynamic case, i.e., positive in the upper plane and negative in the lower plane.

We now construct an $\alpha_{\phi\phi}$ profile based on the results of \cite{Gressel2010}, whose overall magnitude follows that of a stratified rotating turbulence,
\beq
\alpha_{\phi\phi}^{(0)}
\simeq-l^2\bm\Omega\cdot\del\ln\rho u^2
=-l^2\Omega\partial_z\ln\rho u^2
=\frac{\gamma z}{h}\alphaSS\cs,
\label{eqn:alpha330}
\eeq
where the superscript $(0)$ denotes the un-quenched value, and we have used $\partial_z\ln\rho u^2\simeq \partial_z\ln\rho\cs^2
=\partial_z\ln P=\partial_z \ln \rho^\gamma=-\gamma\partial_z(z^2/2h^2)$.
This expression only differs from the more sophisticated calculation of \cite{RuedigerPipin2000} by a factor of $\zeta/5$, where $\zeta=\abra{b^2}/\mu_0\abra{\rho u^2}$ is the ratio between the turbulent magnetic energy and kinetic energy.
Local shearing-box simulations indeed suggest that $\zeta\simeq\mathcal{O}(5)$,
and hence equation~(\ref{eqn:alpha330}) is a fair estimate.

\begin{figure}
\centering
\includegraphics[width=0.8\columnwidth]{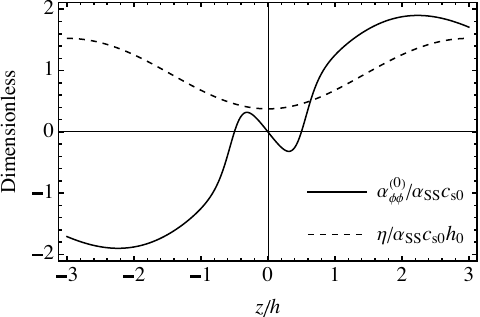}
\caption{The initial vertical profile of the dynamo driver $\alpha$ and turbulent diffusion $\eta$ coefficients.
}
\label{fig:alpha_eta}
\end{figure}

Near the disc mid-plane, the sign reversal of $\alpha_{\phi\phi}$ is captured in an \textit{ad hoc} way through an additional $z$ dependence to equation~(\ref{eqn:alpha330}):
\beq
\alpha_{\phi\phi}^{(0)}=\frac{\gamma z}{h}\alphaSS\cs
\tanh\left(\frac{5|z|}{h}-\frac{5}{2}\right).
\label{eqn:alpha_ao}
\eeq
Note that $\cs$ has both radial (power-law) and vertical (exponential) dependence, which self-consistently gives the full profile of $\alpha_{\phi\phi}^{(0)}$.
The initial vertical profile of $\alpha_{\phi\phi}^{(0)}$ and $\eta$ are plotted in figure~\ref{fig:alpha_eta}, to be compared with the test-field results of \cite{Gressel2010}.

$\alpha_{\phi\phi}^{(0)}$ drives a helical dynamo that feedbacks onto the turbulent EMF, and is nonlinearly quenched.
In previous GRMHD simulations with sub-grid dynamo terms, it is typical to use the parametrized form
\beq
\alpha_{\phi\phi,\text{par}}=\frac{\alpha_{\phi\phi}^{(0)}}{1+B^2/B_\text{st}^2},
\label{eqn:a33alg}
\eeq
where $B_\text{st}$ is some target saturation value.
In the work by \cite{MattiaFendt2022}, the effects of different quenching formulae on jet launching and production of flares in the inner disc region has been explored.
In this work, we consider a dynamical quenching model constrained by magnetic helicity conservation, similar to those applied to stellar and galactic dynamo models.

In the dynamical quenching formalism \citep{BlackmanField2002}, 
$\alpha_{ij}$ receives a back-reacting magnetic contribution $\alphaM_{ij}$ whose trace is $\tau\abra{\bm j\cdot\bmb}/\mu_0\rho$, i.e., proportional to the small-scale current helicity density $\abra{\bm j\cdot\bmb}$ where $\bm j$ is the small-scale current helicity.
We assume that $\alphaM_{\phi\phi}$ has the same $z$-dependence as that of $\alpha_{\phi\phi}^{(0)}$, and the full form of $\alpha_{\phi\phi}$ is taken to be
\beq
\alpha_{\phi\phi}=\frac{\gamma z}{h}\tanh\left(\frac{5|z|}{h}-\frac{5}{2}\right)\left(\alphaSS\cs-\alphaM\right),
\eeq
where $\alphaM=\tau\left|\abra{\bm j\cdot\bmb}\right|/3\mu_0\rho$.

To connect $\alphaM$ to the large-scale fields, we consider the small-scale magnetic helicity associated with the current helicity,
\beq
\abra{\bma\cdot\bmb}\simeq k_2^{-2}\abra{\bmj\cdot\bmb},
\label{eqn:abjb}
\eeq
where $\bm a$ is the small-scale vector potential.
If no helicity flux is present, conservation of magnetic helicity is hold locally at scale $\simeq h$,
\beq
\abra{\bma\cdot\bmb}+\bmA\cdot\bmB=0\ \text{for all }t,
\label{eqn:ab+AB}
\eeq
assuming that the dynamo has started with negligible total magnetic helicity.
Here $\bm A$ is the large-scale vector potential.
The right side of equation~(\ref{eqn:ab+AB}) becomes nonzero
if (i) the helicity density is advected by a mean flow, or (ii) a Vishniac-type flux arises due to the inhomogeneities of small-scale fields.
We now consider such two possibilities.

Large-scale outflows like disc winds preferentially carry away large-scale magnetic fields since small-scale fields cannot survive the shredding by the shear flow when they buoyantly arise \citep{BlackmanPessah2009}.
This leads to a deficit on the right side of equation~(\ref{eqn:ab+AB}), which
we parametrize as a certain fraction $\fw$ of the large-scale magnetic helicity,
\beq
\abra{\bma\cdot\bmb}+\bmA\cdot\bmB=-\fw \bmA\cdot\bmB,\ \fw\geq0.
\label{eqn:hel_loss}
\eeq
If $U_\text{w}$ is the typical wind speed leaving the disc and $\Omega^{-1}$ is the typical time scale for helicity loss,
we have the rough estimate that
\beq
\fw\sim\frac{\Omega^{-1}\bm U_\text{w}\cdot\del|\bmA\cdot\bmB|}{|\bmA\cdot\bmB|}
\simeq\frac{\Omega^{-1}U_\text{w}}{h}
\simeq \frac{U_\text{w}}{\cs},
\eeq
i.e., $\fw$ is roughly equal to the Mach number of the disc wind.
In general $U_\text{w}/\cs$ varies radially, and more importantly, depends on the local magnetic field strength.
For strong poloidal magnetic fields whose Alfv\'en speed is larger than the local sound speed, disc winds are magneto-centrifugally driven, whereas for weaker poloidal magnetic fields, disc winds are driven by the magnetic pressure of the toroidal fields.
Focusing on the dynamo processes within the disc, we adopt constant values of $\fw$ in this work.

Helicity flux may also arise from triple correlations of small-scale fields,
as considered by \cite{GopalakrishnanSubramanian2023}.
The derived flux depends on the gradients of $\abra{u^2}$ and $\abra{b^2}$ along the direction of the mean vorticity, i.e., in the $z$ direction.
Using equation~(18) in \cite{GopalakrishnanSubramanian2023}, we find that the contribution from such terms is $\Delta\abra{\bm a\cdot\bmb}\simeq\alphaSS h \abra{b^2}(0.08\zeta-0.04)$ in the upper disc plane during one Keplerian time scale.
For later convenience we also write it as a fraction of the large-scale magnetic helicity, $f_\text{f}\bmA\cdot\bmB$, with
\beq
f_\text{f}=\frac{\Delta\abra{\bm a\cdot\bmb}}{\bm A\cdot\bmB}
=\alphaSS^2\zeta(0.04\zeta-0.02)\betaL,
\label{eqn:ff}
\eeq
where we have used $\abra{b^2}/{B^2}=(\abra{b^2}/{\rho u^2})(\rho u^2/B^2)=\alphaSS\zeta\betaL/2$, and
\beq
\betaL=\frac{2\mu_0 P}{B^2}
\eeq
is the ratio between the thermal pressure and the magnetic pressure from the large-scale fields.
Note that equation~(\ref{eqn:ff}) gives a positive $\ff$ and hence a positive contribution to $\abra{\bma\cdot\bmb}$, in contrast to the thin-disc estimation of \cite{GopalakrishnanSubramanian2023}, due to the difference between the models.

Collecting all the terms, we have
\begin{align}
&\left|\abra{\bm j\cdot\bmb}\right|\sim k_2^2\left|\abra{\bm a\cdot\bmb}\right|\notag\\
&\sim (1+\fw-\ff)k_2^2\left|\bm A\cdot\bm B\right|\sim (1+\fw-\ff)k_2^2/k_1 B^2.
\end{align}
Denoting $f=1+\fw-\ff$, it follows that
\beq
\frac{\tau}{3\rho}\left|\abra{\bm j\cdot\bm b}\right|\simeq f\frac{\tau}{3\rho}\frac{k_2^2}{k_1}B^2
=\frac{f}{3\rho}\frac{h}{ul}B^2=\frac{2f}{3\alphaSS\betaL}\cs.
\label{eqn:jb}
\eeq
Combining both kinetic and magnetic contributions, we have
\beq
\alpha_{\phi\phi}=\frac{\gamma z}{h}
\tanh\left(\frac{5|z|}{h}-\frac{5}{2}\right)\alpha(\betaL),
\label{eqn:alphaij_ao}
\eeq
where
\beq
\alpha(\betaL)=\left[\alphaSS-\frac{2f}{3\alphaSS\betaL}\right]\cs.
\label{eqn:alpha_betaL}
\eeq

Note that equation~(\ref{eqn:alphaij_ao}) is only valid within the disc where the gas is sufficiently turbulent.
In the numerical solutions, we do not allow $\alpha(\betaL)$ to become negative, i.e., $\alpha_{\phi\phi}$ does not pass zero.
Changes of signs for $\alpha_{\phi\phi}$ mostly happen when a helical magnetic loop buoyantly rises into the corona, and then ``untwists'' due to the inverse transfer of magnetic helicity.

Given the full expression of $\alpha_{\phi\phi}$, the dynamo saturation level can be estimated from the dynamo number,
i.e., the ratio between the products of the constructive and the destructive coefficients in the dynamo equation.
As order-of-magnitude estimates, we simply use $\alpha_{\phi\phi}\simeq\alpha(\betaL)$, ignoring its $z$-dependence, and $f=1$ for the moment.
The dynamo number for this $\alpha\Omega$ dynamo is
\beq
D(\betaL)=\frac{\alpha(\betaL)\Omega}{\eta^2 k_1^3}.
\eeq
At dynamo saturation we have $D(\beta_\text{L,st})=1$, yielding
\beq
\beta_\text{L,st}=\frac{2}{3\alphaSS^2(1-\alphaSS)}.
\label{eqn:betaLst_a2o}
\eeq
Equation~(\ref{eqn:betaLst_a2o}) gives $\beta_\text{L,st}\simeq 74$ if $\alphaSS=0.1$,
and $\beta_\text{L,st}\simeq 11$ if $\alphaSS=0.3$.
These values are only $\sim 2$ times larger than those derived from the exact dispersion relation.

A few remarks are made regarding connections to previous works:
(i) \cite{Pudritz1981} derived a critical value $\beta_\text{L,P81}=4/3c_u\alpha$ (where $c_u\lesssim1/2$) based on energy balance between the magnetic and kinetic turbulent energy.
Our estimation from the mean-field dynamo theory, equation~(\ref{eqn:betaLst_a2o}), is $\sim3$ times larger than $\beta_\text{L,P81}$ if $\alphaSS=0.1$.
However, as we shall see in section~\ref{sec:results}, the actual saturation value of $\betaL$ is much larger than $\betaLst$, possibly due to the more complicated profile of $\alpha_{\phi\phi}$ in the disc than that in the simplified scenario considered above.
(ii) The energy equipartition between the large-scale field and the turbulent kinetic energy gives
\beq
\betaLeq\simeq\frac{2P}{\rho u^2}=\frac{2\cs^2}{u^2}=\frac{2}{\alphaSS},
\label{eqn:betaLeq}
\eeq
Since
\beq
\frac{\betaLst}{\betaLeq}=\frac{1}{3\alphaSS(1-\alphaSS)},
\eeq
the saturated magnetic field is a few times smaller than its energy-equipartition value.
(iii) The quenching formula~(\ref{eqn:alphaij_ao}) combined
with equation~(\ref{eqn:betaLst_a2o}) yields a faster quenching than the more often used parametrized quenching~(\ref{eqn:a33alg}).
For equation~(\ref{eqn:alphaij_ao}), the quenched value of $\alpha$ at dynamo saturation is
\beq
\frac{\alpha(\betaLst)}{\alpha(\infty)}
=\frac{\alphaSS}{1-\alphaSS}
\sim\mathcal{O}(0.1)
\eeq
times its kinematic value,
whereas the typically used form gives
\beq
\frac{\alpha_{\phi\phi,\text{par}}(\betaLst)}
{\alpha_{\phi\phi,\text{par}}(\betaL)}
=\frac{1}{2},
\eeq
i.e., only $50\%$ quenching when $\betaL=\betaLst$.
Hence the latter formula will lead to a lower saturated value of $\betaL$ than $\betaLst$.

\subsection{The incoherent $\alpha$ model}
In this and the next subsections, we introduce the two nonhelical large-scale dynamo mechanisms that originate from forced shearing-box simulations \citep{Brandenburg2005,Yousef+2008}, and couple them to the thin-disc model.
In contrast to the conventional $\alpha$ mechanism, these two mechanisms do not require a net kinetic helicity in the flow, but only the combination of turbulence and a shearing flow.

The IAE relies on the fluctuations of the kinetic helicity and therefore a fluctuating $\alpha$ term with zero mean.
The possibility of dynamo driven by fluctuating helicity was first raise by \cite{Kraichnan1976} and \cite{Moffatt1978} in non-shearing turbulence,
and numerically by \cite{VishniacBrandenburg1997}.
More recently, \cite{SridharSingh2014} and \cite{Jingade+2018} showed that in the presence of shear, such helicity fluctuations can be recast into an effective anisotropic turbulent magnetic diffusivity tensor which is capable of a dynamo action.

In dynamos driven by IAE, the turbulent EMF takes the form
\beq
\bm\emf=\alpha\bmB-\eta\bm J,
\label{eqn:emf_fa}
\eeq
where the pseudo-scalar $\alpha$ is inhomogeneous in space and fluctuates in time.
We denote the values of $\alpha$ in the un-quenched regime by $\alpha_0$.

The $\alpha$ fluctuations are assumed to happen on meso-scales, i.e., at spatio-temporal scales larger than those of the turbulent flow but still smaller than the mean-field scale.
We parameterise such scales using $l_\alpha=ml$ for the length scale and $\tau_\alpha=m\Omega^{-1}$ for the time scale,
where $m>1$ is a model parameter.
During the kinematic regime, $m$ is solely a property of the turbulent flow, but unfortunately it is so far unclear how its value can be calculated theoretically.
An attempt to determine it in simulations is on-going (Zhou and Jingade, \emph{in prep.}), and preliminary results indicate that $m$ is $\lesssim\mathcal{O}(5)$ and scales weakly with the shear rate $S$ as $(S\tau)^{0.3}$.
In our thin-disc model, since $S=3\Omega/2$ and hence $S\tau=3/2$ is a constant, we shall take a constant and uniform $m$ throughout the disc.

Another necessary ingredient of the model is what probability distribution function (PDF) the $\alpha$ fluctuations follow.
\cite{JingadeSingh2021} have used a uniform PDF in their renovating-flow model, where $\alpha_0\in[-\alpha_\text{max},\alpha_\text{max}]$ and $\alpha_\text{max}=u/3$ is the fully helical value.
Under such conditions they found that $m\geq3$ is needed for a large-scale dynamo.
However, the actual PDF of $\alpha$ fluctuations in realistic turbulence is not clear.
In forced shearing turbulence, \cite{Brandenburg+2008} found $\alpha_0$ to follow a Gaussian distribution, $\alpha_0\sim\mathcal{N}(0,0.2\eta k_2)=\mathcal{N}(0,0.2\alphaSS^{1/2}\cs)$, which has a variance $25$ times smaller than that from a uniform PDF, and hence less capable of driving a dynamo.
On the other hand, the PDF of $\alpha$ fluctuations in MRI turbulence has not been determined in simulations yet.
In this work, we explore both uniform and Gaussian PDFs for $\alpha_0$ to check their dynamo capability in a disc geometry.

Following \cite{Kraichnan1976}, the effective EMF for the incoherent $\alpha$ dynamo is
\beq
\emf=-\left(\sigma^2\alphaSS\cs^2\tau_\alpha-\beta\right)\bm J,
\eeq
where $\sigma^2=\abra{\alpha^2}/\alphaSS\cs^2$ is the normalized variance of the $\alpha$ fluctuations determined by the PDF.
The dynamo growth rate in the kinematic regime is
\beq
\gamma_\text{IA}\Omega^{-1}=\alphaSS(m\sigma_0^2-1),
\eeq
with $\sigma_0^2=\abra{\alpha^2_0}/\alphaSS\cs^2$.

The nonlinear saturation of the incoherent $\alpha$ dynamo is so far unclear.
Numerical simulations of unstratified or stratified MRI turbulence suggest that the large-scale magnetic field strength typically saturate at $\mathcal{O}(1)$ times the energy-equipartition value \citep{Brandenburg+1995,Davis+2010,Shi+2016,Guilet+2022}, $\betaLeq=2/\alphaSS$ [see equation~(\ref{eqn:betaLeq})].
To achieve this, we may use a parameterised quenching formula
\beq
\alpha=\frac{\alpha_0}{1+q\betaLeq/\betaL},
\eeq
and correspondingly $\sigma=\sigma_0/(1+q\betaLeq/\betaL)$.
The factor $q$ is to be determined by the condition $\gamma_\text{IA}(\betaL=\betaLeq)=0$, giving $q=\sqrt{m}\sigma_0-1$.
Collecting all the terms we have
\beq
\alpha=\alpha_0\left[
1+\left(\sqrt{m}\sigma_0-1\right)\frac{2}{\alphaSS\betaL}\right]^{-1}.
\label{eqn:alphatot_fa}
\eeq

\begin{figure}
\includegraphics[width=\columnwidth]{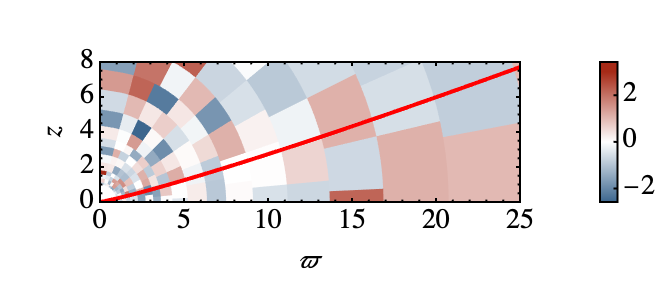}
\caption{An example of the constructed meso-scale cells ($m=3$) of $\alpha$ fluctuations assigned with random values.
The thick red curve indicates the disc-corona boundary at $\pi/2-\theta=2\epsilon$.}
\label{fig:alpha_cells}
\end{figure}

To implement meso-scale fluctuations of $\alpha$ as a sub-grid model,
we first divide the simulation domain into cells of scale $l_\alpha=ml=m\alphaSS^{1/2}h$, and within each cell the value of $\alpha_0$ is uniform but randomly drawn from a given PDF at time intervals roughly equal to $\tau_\alpha=m\Omega^{-1}$\footnote{%
This is done by generating a random number $\texttt{rand}\in[0,1]$ for each cell at each time step,
and if $\texttt{rand}<\texttt{dt}/\tau_\alpha$ with $\texttt{dt}$ being the time step,
the cell value is updated.}.
An illustration of the coherent cells of $\alpha$ is in figure~\ref{fig:alpha_cells}.

\subsection{The shear-current effect}
The second nonhelical large-scale dynamo mechanism is the shear-current effect \citep{RogachevskiiKleeorin2003,RogachevskiiKleeorin2004,SquireBhattacharjee2015},
where the $\alpha_{ij}$ tensor vanishes but the diffusivity tensor has non-zero off-diagonal components,
\beq
\emf_i=-\eta_{ij}J_j.
\label{eqn:emf_sc}
\eeq
For Keplerian flows, \cite{ZhouBlackman2021shear} used the spectral-$\tau$ closure and derived that
\beq
\eta_{r\phi,0}=(0.63-0.26\zeta)\Sh\eta,\ 
\eta_{\phi r,0}=(-0.03-0.124\zeta)\Sh\eta,
\label{eqn:eta_sc}
\eeq
where $\zeta=\abra{b^2}/\mu_0\rho\abra{u^2}$, $\Sh=S\tau=3/2$ is the dimensionless shear number, and the subscripts $0$ indicate that they are kinematic values.
The dynamo growth rate is
\beq
\gamma_\text{SC}=\left(
\eta_{r\phi}\eta_{\phi r}k_1^4-S\eta_{\phi r}k_1^2\right)^{1/2}
-\frac{\eta_{rr}+\eta_{\phi\phi}}{2}k_1^2.
\label{eqn:-gamma_sc}
\eeq
Using the specific forms equation~(\ref{eqn:eta_sc}), the normalized dynamo growth rate is found to be
\begin{align}
&\gamma_\text{SC}\Omega^{-1}\simeq -\alphaSS\notag\\
&+0.27\sqrt{\alphaSS(0.24+\zeta)(3.85-2.42\alphaSS+\alphaSS\zeta)}.
\label{eqn:gamma_sc}
\end{align}

We now consider a quenching prescription for the shear-current effect.
Similar to the incoherent-$\alpha$ model, we propose the parametrized form
\beq
\eta_{r\phi,\phi r}=\frac{\eta_{r\phi,\phi r,0}}{1+q\betaLeq/\betaL},
\eeq
and $q$ is to be determined by requiring that the dynamo growth rate vanishes when $\betaL=\betaLeq$.
At $\alphaSS=0.3$, this gives approximately
\beq
q\simeq \zeta-0.89.
\eeq
Hence an \textit{ad hoc} quenching prescription can be
\beq
\eta_{r\phi,\phi r}=\eta_{r\phi,\phi r,0}\left[1+\frac{2\zeta-1.78}{\alphaSS\betaL}\right]^{-1}.
\label{eqn:Sh_sc}
\eeq

\section{Numerical setup}
\label{sec:simulations}
In this section we introduce the setup in the publicly available \textsc{Pencil Code} \citep{JOSS2021} to find numerical solutions of the dynamo models derived in the last section.
The physical quantities are independent of the azimuthal angle, but vectors still have finite $\phi$ components, i.e., the simulations are 2.5-dimensional.

We do not resolve the turbulent-scale fields, but only incorporate them as sub-grid physics including turbulent diffusion of mass, velocity and magnetic fields, and the dynamo coefficients.
In this sense, the simulations shall be understood as numerical solutions of mean-field models, rather than DNS of accretion discs.

We consider azimuthally averaged MHD equations with a central gravitational source.
Furthermore, since viscous and resistive dissipations are just energy transfers between the large- and small-scale fields, we assume such processes do not change the mean-field entropy and thereby consider a polytropic gas.
The equations to be solved are
\begin{align}
\partial_t\rho+\del\cdot(\rho \bmU)=&\chi\nabla^2\rho,\\
\partial_t\bmU+\bmU\cdot\del\bmU=&\frac{1}{\rho}\left[-\del P
+\bm J\times\bm B+\del\cdot\left(2\nu\rho\bm S\right)\right]\notag\\
&-\frac{GM\hat{\bm r}}{r^2},\label{eqn:dudt}\\
\partial_t\bm A=&\bmU\times\bm B+\bm\emf.
\label{eqn:induction_A}
\end{align}
Here $\chi$ is the turbulent mass diffusion coefficient taken to be equal to $\nu$,
and $S_{ij}=(\partial_i U_j+\partial_j U_i)/2-\delta_{ij}\del\cdot\bm U/3$ is the rate-of-strain tensor.

To ensure that shocks are properly resolved, we add a shock-capturing diffusion term on the right-hand side of equation~(\ref{eqn:dudt}),
which is a bulk viscosity of the form $f_\text{shock}=\rho^{-1}\del\left(\xi\rho\del\cdot\bm U\right)$,
and
\beq
\xi=\left\{\begin{aligned}
&\text{max}_5\left|\del\cdot\bm U\right|
\left[\text{min}\left(\Delta r,r\Delta\theta\right)\right]^2,
&\text{if }\del\cdot\bm U<0\\
&0, &\text{otherwise}
\end{aligned}\right\}.
\eeq
Here $\text{max}_5$ means the maximum value in the neighboring five mesh points, and $\Delta r$ and $\Delta\theta$ are the mesh sizes in the radial and polar directions, respectively.
Hence the shock diffusion is the strongest for the strongly converging flows.

The simulations are carried out in spherical coordinates,
with the simulation domain being $r\in[r_0,100r_0]$ and $\theta\in[0,\pi]$.
We use a resolution of $N_r\times N_\theta=384\times256$ and static mesh refinement in the radial direction, i.e.,
when moving towards the disc inner region, we double the resolution inside $16r_0$, and further double it inside $8r_0$.

A summary of the runs is in table~\ref{tab:runs}.
For all the runs, the magnetic field is only initiated after the accretion flow has reached a steady state at $t=4000\Omega_0^{-1}$.

\begin{table}
\centering
\caption{Summary of the simulations.
The prefix in the run names \texttt{AO} refers to $\alpha\Omega$ models,
\texttt{IA} refers to incoherent $\alpha$ models,
and \texttt{SC} refers to shear-current models.
The last column labels the resulting dynamo modes after saturation:
D for a dipole mode, Q for a quadruple mode, and
F for strongly fluctuating polarization.}
\label{tab:runs}
\begin{tabular}{cccccccc}
\hline
Run & Quenching & $\fw$ & $\ff$ & PDF & $m$ & $\zeta$ & Dynamo?\\
\hline
\texttt{AO1}  & (\ref{eqn:alphaij_ao}) & 0 & N &/& / & 5 & D\\
\texttt{AO2}  & (\ref{eqn:a33alg}) & 0 & N &/& / & 5 & Q\\
\texttt{AO3}  & (\ref{eqn:alphaij_ao}) & 0 & Y &/& / & 5 & D\\
\texttt{AO4}  & (\ref{eqn:alphaij_ao}) & 1 & Y &/& / & 5 & D\\
\texttt{IA1}   & (\ref{eqn:alphatot_fa}) & / & / & Uniform & 3 & / & F\\
\texttt{SC1}   & (\ref{eqn:Sh_sc}) & / & / & / & / & 5 & Q\\
\hline
\end{tabular}
\end{table}

\subsection{Boundary conditions}
The boundary conditions mostly follow those of \cite{MattiaFendt2022}, with a few adaptions for the \textsc{Pencil Code} since the latter uses a finite-difference scheme and evolves the vector potential rather than the magnetic field.

In the radial direction, $U_r$ is extrapolated in a power law with index $-3/2$ at the inner boundary, and uses outflow boundary condition at the outer boundary.
$U_\theta$ uses free boundary condition with vanishing second-order derivative at the inner boundary, and is $\propto r^0$ at the outer boundary.
$U_\phi$ is kept at $98\%$ times the local Keplerian value at both radial boundaries.
For the density field and all the three components of the magnetic vector potential, free boundary condition is used for both boundaries.

In the polar direction, the usual symmetric or anti-symmetric boundary condition is used.

\subsection{Initial conditions}
The initial density follows
\beq
\rho(t=0)=\rho_0 r^{p_\rho} e^{-z^2/2h^2},
\eeq
where $p_\rho=-15/8$ for a standard Shakura-Sunyaev thin disc.
We also impose a density floor
\beq
\rho_\text{floor}=\rho_{\text{floor},0}(r/r_0)^{p_\rho},
\eeq
where $\rho_{\text{floor},0}=10^{-3}\rho_0$.

Within the disc, the initial azimuthal velocity is $98\%$ of the Keplerian value, while in the corona the gas has zero initial velocity.

The initial vector potential is $\bm A(t=0)\propto-\theta/r\hat{\bm r}$,
which yields a purely toroidal initial field $B_\phi(t=0)\propto r^{-2}$.
The initial disc magnetization is weak, with $\betaL>10^9$.

\subsection{Model parameters}
For all the simulations in this work, we use $\cso=0.1$ and $\gamma=1.4$ for a cold disc,
and $GM=\rho_0=1$.

The Shakura-Sunyaev prescription gives $\nu=\alphaSS\cs h\propto \varpi^{p_\rho(\gamma-1)+3/2}=\varpi^{-3/4}$.
However, a steeper profile, $\nu\propto \varpi^{-1}$, is actually used to
ensure a stable accretion flow.
Hence the turbulent viscosity is kept fixed in space and time and given by
\beq
\nu=\alphaSS\cso h_0\left(\frac{\varpi}{\varpi_0}\right)^{-1}.
\eeq

On the other hand, the turbulent resistivity is determined by the local gas properties through
\beq
\eta=\alpha\cs h=\alphaSS\frac{\cs^2}{\Omega}.
\eeq
In the corona region where $|\theta-\pi/2|\geq 2\epsilon$,
we use fixed values $\eta=\nu$ to prevent magnetic field lines from accumulating and eventually crashing the simulation.

For all the dynamo models, we restrict the dynamo-driving coefficients (i.e., all the components of $\alpha_{ij}$ and the off-diagonal components of $\eta_{ij}$) inside the disc by applying to them an additional spatial profile
\beq
f_\text{disc}(z)=\frac{1}{2}+\frac{1}{2}\tanh(4-2|z|/h),
\eeq
which rapidly damps the dynamo drivers beyond two disc scale heights.

\section{Results}
\label{sec:results}
In this section we report the numerical solutions for each dynamo models with different choices of parameters.
We first give a cross comparison between the fiducial runs for each dynamo models, and then explore the $\alpha\Omega$ and the IAE models in more detail.
A summary of the resulting magnetic field configurations is listed in the last column of table~\ref{tab:runs}.

\subsection{Cross comparison among the fiducial runs}
We first  compare results from the fiducial runs from each dynamo model,
runs~\texttt{AO1}, \texttt{IA1}, and \texttt{SC1}.
In the fiducial $\alpha\Omega$ model \run{AO1}, no helicity flux is considered and hence $f=1$.
For the incoherent $\alpha$ model \run{IA1}, we use $m=3$ and a uniform PDF for $\alpha$.
For the shear-current model \run{SC1}, we use $\zeta=5$.

\begin{figure}
\centering
\includegraphics[width=0.9\columnwidth]{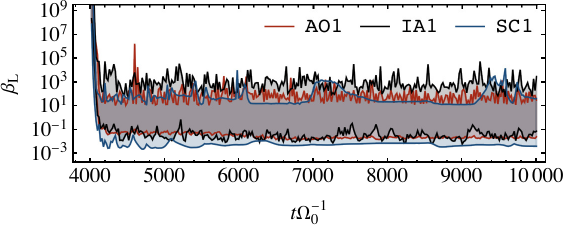}
\caption{Maximal and minimal values of $\betaL$ in the inner disc region for the fiducial runs.}
\label{fig:betad}
\end{figure}

The time evolution of the maximum and minimum $\betaL$ values in the region $r\in[\rin,2\rin]$ and $\theta\in[\pi/2-\epsilon,\pi/2+\epsilon]$ is plotted in figure~\ref{fig:betad} for each model.
All the three models share a similar kinematic growth rate.
Due to the random nature of the the incoherent-$\alpha$ model, \run{IA1} yields strongly fluctuating fields with the weakest strength.
In contrast, the shear-current model yields the strongest disc magnetization and the least fluctuations.
The mean values of $\betaL$ are roughly $100$, $100$, and $10$ for runs~\texttt{AO1}, \texttt{IA1}, and \texttt{SC1}, respectively.

\begin{figure}
\centering
\includegraphics[width=0.9\columnwidth]{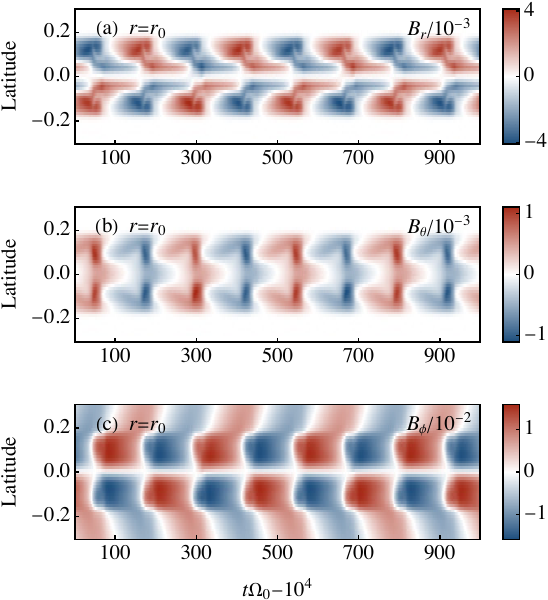}
\caption{Space-time diagram for the three components of the magnetic field at $r=r_0$ for the fiducial $\alpha\Omega$ dynamo, \run{AO1}.}
\label{fig:ao_l3_vid_butt}
\end{figure}

\begin{figure}
\centering
\includegraphics[width=0.7\columnwidth]{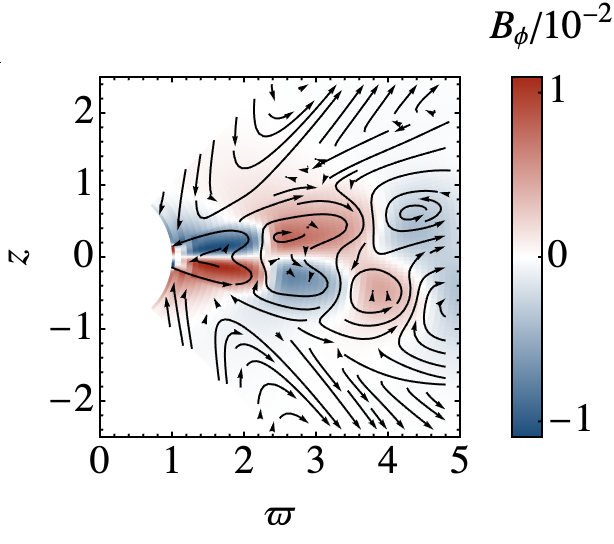}
\caption{The snapshot at $t=10550\Omega_0^{-1}$ for the poloidal fields (solid curves with arrows) and the toroidal field (color),
for \run{AO1}.}
\label{fig:ao_l3_vid_snap}
\end{figure}

In figure~\ref{fig:ao_l3_vid_butt}, we show the space-time diagram of the $\alpha\Omega$ model in its steady state.
The azimuthal component $B_\phi$ reproduces the butterfly diagrams seen in previous local \citep{Gressel2010} and global \citep{HoggReynolds2018} simulations, where the dynamo waves migrates away from the disc mid-plane.
However, the radial component $B_r$ in the high latitude region, $z/h\geq 2$, is rather weak and does not follow the pattern seen in \cite{Gressel2010}.
This might result from that we have restrict the dynamo terms inside the disc.
A snapshot of the field configuration is shown in figure~\ref{fig:ao_l3_vid_snap},
where the coherent scale of $B_\phi$ is $\simeq h$ in the vertical direction, and $\simeq 10h$ in the radial direction.
The alternating field polarity is in general agreement with global simulations of disc dynamos by \cite{HoggReynolds2016}.

Overall, the amplified large-scale magnetic field resembles a dipolar configuration with a cycle period of $\sim 40$ orbital time scale at $r_0$, in agreement with previous global simulation results \citep{HoggReynolds2018}.
Note that this is a few times longer than the estimate from local periodic-box dispersion relations \citep{GresselPessah2015} and also previous sub-grid GRMHD simulations without $\alpha$ quenching \citep{Bugli+2014}, which is about $\sim10$ times the orbital time scale.

\begin{figure}
\centering
\includegraphics[width=0.9\columnwidth]{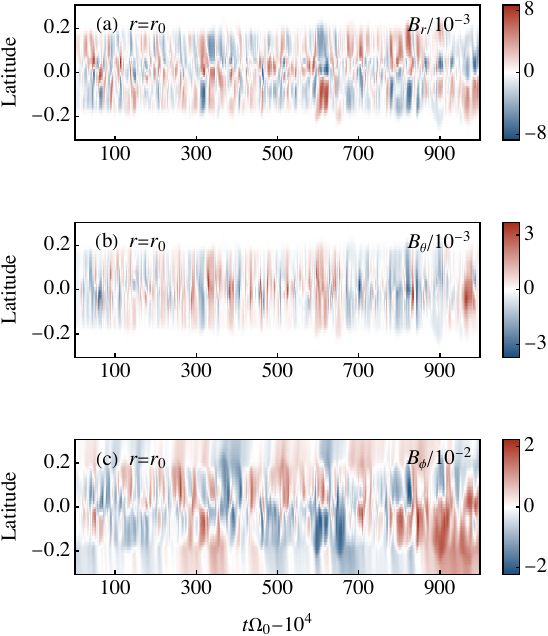}
\caption{Space-time diagram for the three components of the magnetic field at $r=r_0$ for the fiducial IAE dynamo, \run{IA1}.}
\label{fig:fa_m3uf_vid_butt}
\end{figure}

\begin{figure}
\centering
\includegraphics[width=0.7\columnwidth]{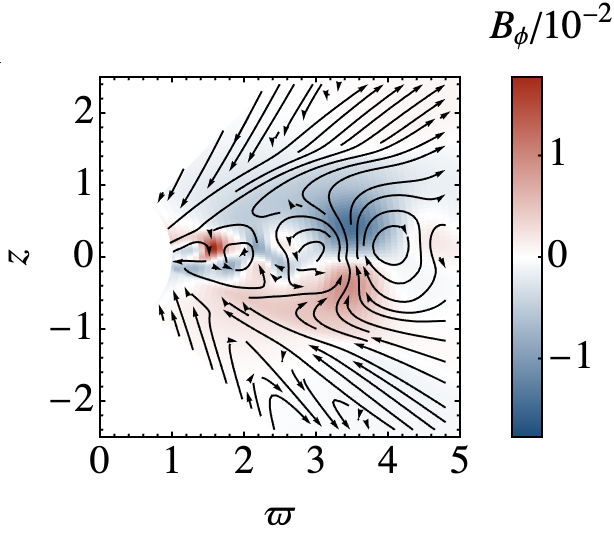}
\caption{The snapshot at $t=10550\Omega_0^{-1}$ for the poloidal fields (solid curves with arrows) and the toroidal field (color),
for \run{IA1}.}
\label{fig:fa_m3uf_vid_snap}
\end{figure}

\begin{figure}
\centering
\includegraphics[width=0.9\columnwidth]{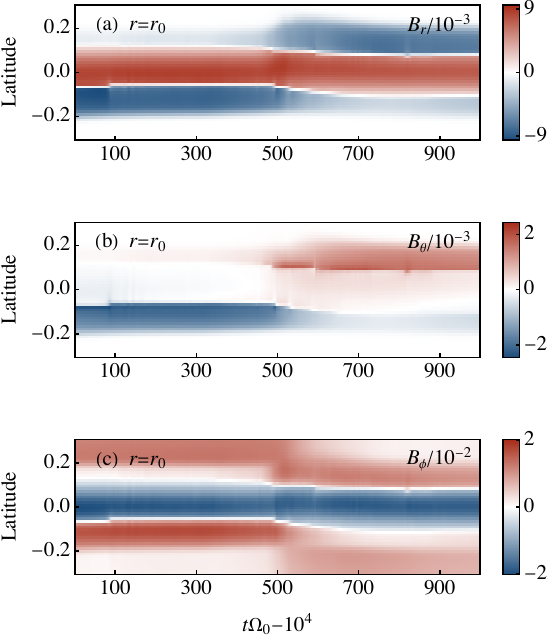}
\caption{Space-time diagram for the three components of the magnetic field at $r=r_0$ for the fiducial SCE dynamo, \run{SC1}.}
\label{fig:sc_xi5_vid_butt}
\end{figure}

\begin{figure}
\centering
\includegraphics[width=0.7\columnwidth]{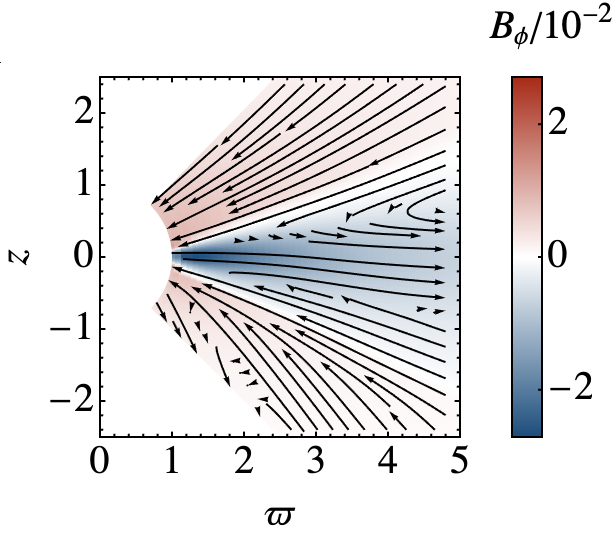}
\caption{The snapshot at $t=10550\Omega_0^{-1}$ for the poloidal fields (solid curves with arrows) and the toroidal field (color),
for \run{SC1}.}
\label{fig:sc_xi5_vid_snap}
\end{figure}

In figures~\ref{fig:fa_m3uf_vid_butt} and \ref{fig:fa_m3uf_vid_snap}, we show the space-time diagrams and a snapshot of field configuration for the fiducial incoherent-$\alpha$ model, run~\texttt{IA1}.
Although the magnetic field is somewhat more coherent in space than that in \run{AO1}, the incoherent-$\alpha$ model has much more temporal fluctuations.
In contrast, figures~\ref{fig:sc_xi5_vid_butt} and \ref{fig:sc_xi5_vid_snap} present the results for the shear-current model \run{SC1}, and show the most coherent field structure in both space and time.
For both of the nonhelical models, they do not present regular dynamo cycles, in contrast to global simulations.
We hence conclude that even though they may co-exist with the helical $\alpha$ effect in discs,
they are not likely the dominant driver.

\subsection{Dynamical versus parametrized quenching for $\alpha$}
We now compare the fiducial $\alpha\Omega$ model with \run{AO2} which employs the often used parametrized quenching prescription, equation~(\ref{eqn:a33alg}).
Both runs have no helicity flux, i.e., $f=1$.

\begin{figure}
\centering
\includegraphics[width=0.9\columnwidth]{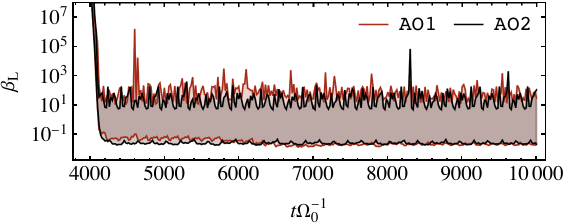}
\caption{Average values of $\betaL$ in the disc region for the $\alpha\Omega$ models with different quenching prescriptions, with $\alphaSS=0.3$.}
\label{fig:ao12_betad}
\end{figure}

\begin{figure}
\centering
\includegraphics[width=0.9\columnwidth]{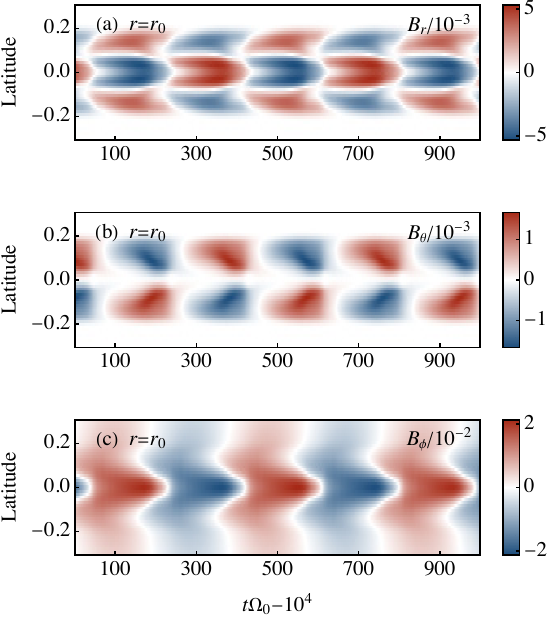}\\
\caption{Space-time diagram for the three components of the magnetic field at $r=r_0$ for the $\alpha\Omega$ dynamo with parameterised. quenching formula (\ref{eqn:a33alg}), \run{AO2}.}
\label{fig:ao_l3_alg_vid_butt}
\end{figure}

In figure~\ref{fig:ao12_betad} we compare the disc magnetization for the two models.
As expected, the two models share similar kinematic growth rates,
but \run{AO2} produces slightly stronger magnetization, particularly at the dynamo minima.
In figure~\ref{fig:ao_l3_alg_vid_butt} we show the space-time diagrams for \run{AO2}.
The cycle period is about $60$ times $2\pi/\Omega_0$, longer than the dynamically quenched case.

Most surprisingly, the parametrized quenching formula produces a quadruple rather than dipole configuration.
In shearing-box simulations, \cite{Brandenburg1998} has suggested that using vertical or perfect-conductor boundary conditions for the magnetic field in the vertical direction determines the parity and cycle period of the dominant large-scale mode in the saturated state.
In our cases, we do not have closed boundaries in the vertical direction, but it is still possible that the quenching formula yields an effective boundary condition at the disc-corona surface.

\subsection{Effects of helicity fluxes}
\begin{figure}
\centering
\includegraphics[width=0.9\columnwidth]{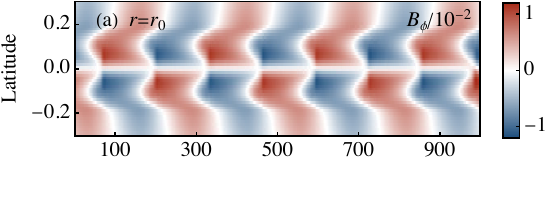}\\
\vspace{-0.3cm}
\includegraphics[width=0.9\columnwidth]{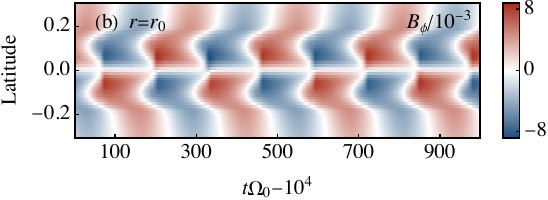}\\
\caption{Space-time diagrams for $B_\phi$ at $r=r_0$ for (a) \run{AO3} and (b) \run{AO4},
both having the new-Vishniac flux but only the latter has the wind loss term with $\fw=1$.}
\label{fig:ao_flux_bb3}
\end{figure}

The effects of helicity fluxes are explored in runs \texttt{AO3} and \texttt{AO4}, both having the helicity flux term but the former uses $\fw=0$ while the latter uses $\fw=1$.
The space-time diagrams for the $B_\phi$ component in these two runs are shown in figure~\ref{fig:ao_flux_bb3}.
Comparing the results of runs~\texttt{AO1} [figure~\ref{fig:ao_l3_vid_butt}(c)], \texttt{AO3} [figure~\ref{fig:ao_flux_bb3}(a)], and \texttt{AO4} [figure~\ref{fig:ao_flux_bb3}(b)], we see that once reaching the steady state,
the flux terms (either Vishniac flux or wind loss) do not have a noticeable impact on the cycle period,  but only slightly change dynamo wave patterns and a decrease in the field strength (a factor of $\lesssim2.5$).

\subsection{Efficiency of the incoherent-$\alpha$ model}
Finally we analyze the dynamo capability of the IAE when using different models of $\alpha$ fluctuations.
Both the coherent scale and the PDF of $\alpha$ fluctuations turn out to be critical.
More specifically, we explore the results with (i) different values of $m$, and (ii) different PDFs.
Future MRI simulations are needed to determined what model (i.e. coherent scale and PDF) best fit realistic turbulence.

\begin{figure}
\centering
\includegraphics[width=0.9\columnwidth]{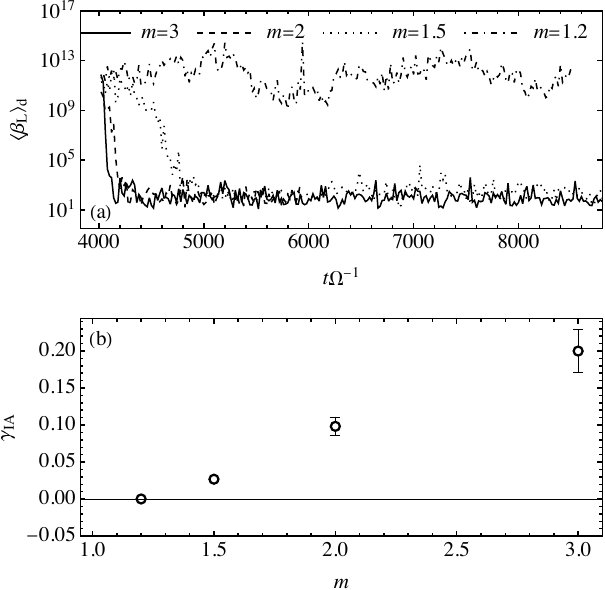}
\caption{Varying $m=l_\alpha/l$ in \run{IA1}.
(a) Time evolution of the mean $\betaL$ in the inner disc.
(b) The kinematic growth rates with varying $m$.}
\label{fig:m_vs_gamma_uf}
\end{figure}

In figure~\ref{fig:m_vs_gamma_uf} we plot the time evolution of the mean $\betaL$ in the region $r\in[\rin,2\rin]$ and $\theta\in[\pi/2-\epsilon,\pi/2+\epsilon]$ in panel (a),
and their fitted kinematic growth rates in panel (b),
by varying the value of $m$ in \run{IA1} but keeping the PDF as a uniform one,
$\alpha_0\sim[-\alpha_\text{max},\alpha_\text{max}]$.
The critical value of $m$ that allows for a dynamo is $m\simeq 1.2$.
This is in general compatible with \cite{Jingade+2018}, who concluded that $m\geq3$ is necessary for a dynamo in their renovating-flow models in a periodic box, rather than in a thin disc.

A uniform PDF produces strong and weak $\alpha$ fluctuations with equal probability, but it is perhaps more plausible that weak to medium strength fluctuations appear more often in realistic turbulence.
In forced shearing-box turbulence, \cite{Brandenburg+2008} found the $\alpha$ fluctuations to follow a Gaussian distribution $\sim\mathcal{N}(0,0.2\alpha_\text{max})$.
However, in such a case we observe no dynamo, i.e., the initial field decays and reaches a steady state where $\betaL\gtrsim10^9$.

\section{Conclusion and discussions}
\label{sec:conclusion}

Implementing sub-grid dynamo terms in low- to medium-resolution GRMHD simulations allow for affordable yet self-consistent amplification of large-scale magnetic fields.
In this work, we formulated both helical and nonhelical dynamo terms in a Shakura-Sunyaev disc, and the derived mean-field models are numerically solved.

For the helical model, we considered several new effects compared to previous implementations: (i) an $\alpha_{ij}$ profile inspired by local MRI simulations, (ii) the dynamical quenching prescription which takes into account magnetic helicity evolution, and (iii) helicity flux terms.
We demonstrate that the resulting large-scale field has a dipole confiuration,
and reproduces the dynamo waves and cycle periods in previous local or global DNS simulations.

For the nonhelical models, we show that they are possible in thin-disc geometries when using optimistic parameters.
However, the resulting dynamo patterns are not consistent with DNS simulations, either with the wrong parity, or with too much fluctuations.
We hence conclude that such dynamos may co-exist with the helical mechanism in accretion discs, but are not the dominant ones.

We do not observe outflows in our simulations, which could be due to
(i) a cold corona which renders the critical radius of Parker wind out of the simulation domain, $r_\text{crit}=GM/2\cs^2\simeq 10^5$, 
(ii) the artificially large magnetic diffusivity in the corona which diffuses out the magnetic fields too quickly before they could relax to meet the Blandford-Payne condition,
and (iii) the lacking of general-relativistic boundary conditions at the inner disc boundary.
A realistic corona is likely heated by the reconnection of magnetic fields, and hence the heating rate depends on $\betaL$ of the coronal fields.
We leave such models in future work.

Disc magnetic fields with alternating polarity may lead to magnetic reconnection near the horizon with the formation of plasmoids and sometimes episodic jets \citep{Parfrey+2015,Mahlmann+2020,Nathanail+2020,Chashkina+2021,Nathanail+2022,Jiang+2023}.
Using sub-grid models, \cite{StepanovsFendt2014} demonstrated such possibilities by switching on and off the dynamo terms by hand.
\cite{FendtGassmann2018} also observed pulsed ejections that are caused by time-varying toroidal fields.
They have noticed that the pulsing time scale is $3-4$ orders of magnitude shorter than those in protostellar jets, but the dynamo cycle period is also rather short, comparable to the local Keplerian time scale.
In our improved model, dynamo cycle periods that are $\sim\mathcal{O}(50)$ Keplerian time scale can be self-consistently produced, and their consequences on jet launching shall be examined.

We have taken $\alphaSS$ as a free parameter throughout, but in a more realistic model it should depend on $\betaL$.
Analytical calculations \citep{RuedigerPipin2000} and numerical simulations \citep{Blackman+2008} found $\alphaSS\propto\betaL^{-1}$ when the saturated value of $\betaL$ is used,
and $\alphaSS\propto\betaL^{-1/2}$ when the initial value of $\betaL$ is used \citep{BegelmanArmitage2023}.
In any case, $\alphaSS$ should increase with decreasing $\betaL$, because a stronger magnetic field taps energy into larger turbulent eddies, and causes stronger turbulent diffusion.
In such cases, some interesting mean-field dynamo features could be (i) a longer kinematic phase because the initial weak field does not produce energetic turbulence
and therefore dynamo effects are weak, and (ii) a weaker quenching effect, because a growing large-scale field leads to stronger turbulence which in turn strengthens dynamo effects, i.e., against dynamo quenching.
The latter effect coupled with accretion dynamics could possibly lead to non-trivial dynamo cycles.
In fact, unifying large-scale dynamo and accretion theories in a mean-field approach has been suggested and investigated by several authors \citep{Pessah+2006,Blackman2012,MondalBhat2023}.

We have considered dynamo models in fully ionized discs where MRI is responsible for the turbulence.
For weakly ionized flows such as young circumstellar discs, different dynamo models are possible \citep{BethuneHenrik2022}.

Finally, we take the chance to highlight that the large-scale dynamo drivers, including the $\alpha$ terms and the shear-current effect,
are all \textit{mean-field} effects, and should only vary on temporal and spatial scales that are larger than the local turbulence scales.
In the current study, we have carried out mean-field simulations which do not resolve the turbulence, and the dynamo coefficients are related to the mean gas density and the mean sound speed.
We propose that a similar implementation should be carried out in GRMHD simulations with sub-grid physics, where the dynamo coefficients should only depend on $\phi$- and time-averaged quantities, rather than the instantaneous values which necessarily include turbulent fluctuations.

\section*{Acknowledgments}
We thank Indu Kalpa Dihingia, Yosuke Mizuno, Eric Blackman and Dong Lai for insightful discussions.
We also thank Axel Brandenburg and Matthias Rheinhardt for the help when setting up the numerical simulations.
HZ acknowledges support from grant 2023M732251 from the China Postdoctoral Science Foundation.
Numerical simulations in this work were carried out 
on the Siyuan-1 cluster supported by the Center for High Performance Computing at Shanghai Jiao Tong University.

\section*{Data availability}
Data and post-processing programs for this work are available on Zenodo at doi:10.5281/zenodo.8379100.

\bibliographystyle{mnras}
\bibliography{refs}

\end{document}